\documentclass[prl,superscriptaddress,amsmath,amssymb,aps,twocolumn,nofootinbib]{revtex4}
\usepackage[dvips]{graphicx}
\usepackage{color}
\usepackage{bm}
\usepackage{mathpazo}
\usepackage{slashed}

\renewcommand\d{\delta}

\renewcommand\l{\lambda}
\renewcommand\r{\rho}

\renewcommand\o{\omega}
\newcommand\e{\epsilon}

\newcommand\m{\mu}
\newcommand\n{\nu}

\newcommand\p{\pi}
\newcommand\h{\theta}
\newcommand\s{\sigma}
\newcommand\f{\phi}

\newcommand\D{\Delta}

\newcommand{\fig}[1]{Fig.~\ref{#1}}
\newcommand{\eq}[1]{Eq.~(\ref{#1})}

\newcommand{\eqs}[2]{Eqs.~(\ref{#1})-(\ref{#2})}

\newcommand\lb{\left(}
\newcommand\rb{\right)}
\newcommand\ls{\left[}
\newcommand\rs{\right]}

\newcommand\ra{\rightarrow}

\newcommand{\non}{\nonumber\\}
\newcommand\pt{\partial}

\newcommand{\diag}{{\rm{diag}}}

\newcommand{\bx}{{\vec x}}

\newcommand{\bk}{{\vec k}}

\newcommand{\bv}{{\vec v}}

\newcommand{\by}{{\vec y}}

\newcommand{\bB}{{\vec B}}
\newcommand{\bE}{{\vec E}}
\newcommand{\bJ}{{\vec J}}

\renewcommand{\part}{{\rm part}}

\renewcommand{\vec}{\boldsymbol}
\newcommand{\be}{\begin{equation}}
\newcommand{\ee}{\end{equation}}
\newcommand{\bear}{\begin{eqnarray}}
\newcommand{\eear}{\end{eqnarray}}
\newcommand{\ba}{\begin{array}}
\newcommand{\ea}{\end{array}}

\begin{document}

\title{On electrodynamics of chiral matter}
\author{\normalsize{Zebin Qiu}}
\affiliation{Physics Department and Center for Particle Physics and Field Theory, Fudan University, Shanghai 200433, China.}
\author{\normalsize{Gaoqing Cao}}
\affiliation{Physics Department and Center for Particle Physics and Field Theory, Fudan University, Shanghai 200433, China.}
\author{\normalsize{Xu-Guang Huang}}
\affiliation{Physics Department and Center for Particle Physics and Field Theory, Fudan University, Shanghai 200433, China.}
\date{\today}

\begin{abstract}
Many-body systems with chiral fermions can exhibit novel transport phenomena that violate parity and time reversal symmetries, such as the chiral magnetic effect, the anomalous Hall effect, and the anomalous generation of charge. Based on the Maxwell-Chern-Simons electrodynamics, we examine some electromagnetic and optical properties of such systems including the electrostatics, the magnetostatics, the propagation of electromagnetic waves, the novel optical effects, etc.
\end{abstract}

\maketitle
\section{Introduction}\label{sec:intro}
Recently, the quantum-anomaly induced transport phenomena in systems with chiral fermions have attracted significant attentions in a wide range of physics, from the Weyl/Dirac semimetals in condensed matter physics~\cite{Miransky:2015ava,Rao:2016,Jia:2016,Burkov:2016,Yan:2016}, the quark-gluon plasma generated in high-energy heavy-ion collisions~\cite{Fukushima:2012vr,Kharzeev:2015kna,Liao:2014ava,Kharzeev:2015znc,Huang:2015oca,Hattori:2016emy}, to the electroweak matter in astrophysics~\cite{Joyce:1997uy,Giovannini:1997eg,Field:1998hi,Boyarsky:2011uy,Boyarsky:2012ex,Yamamoto:2015gzz}. One well-known example of such transport phenomena is the chiral magnetic effect (CME)~\cite{Kharzeev:2007jp,Fukushima:2008xe}, which is the generation of electric current induced by an applied magnetic field in the presence of imbalance between the chemical potentials of right-handed and left-handed fermions,
\begin{eqnarray}
\vec J_{\rm CME} = \frac{e^2}{4\p^2}\D\mu\vec B,
\end{eqnarray}
where $\vec J_{\rm CME}$ is the electric current, $\vec B$ is the applied magnetic field, $e$ is the charge of the fermion, and $\D\m\equiv \m_R-\m_L$ is the imbalance between chemical potentials of right-handed and left-handed fermions which is also called the chiral chemical potential. Another example is the anomalous Hall effect (AHE) in the presence of a separation between the energy-crossing points in momentum space (called Weyl nodes) for right-handed and left-handed fermions characterized by a vector $\D\vec p$~\cite{Haldane:2004zz,Xiao:2009rm,Zyuzin:2012tv,Huang:2015mga,liu:2013},
\begin{eqnarray}
\vec J_{\rm Hall} = \frac{e^2}{4\p^2}\D\vec p\times\vec E,
\end{eqnarray}
where $\vec E$ is the applied electric field. The presence of $\D\vec p$ can lead to another anomalous effect when a magnetic field is present which generates an extra charge density~\cite{Wilczek:1987mv}:
\begin{eqnarray}
J^0_{\rm Anom} = -\frac{e^2}{4\p^2}\D\vec p\cdot\vec B.
\end{eqnarray}
The appearance of the above three quantum-anomaly induced effects changes the electromagnetic response of the chiral matter and turns the electrodynamics of the chiral matter into the following form,
\begin{eqnarray}
\label{csm1}
&&{\vec\nabla}\cdot\bE=J^0-{\vec b}\cdot\bB,\\
\label{csm2}
&&{\vec\nabla}\times\bB-\frac{\pt \bE}{\pt t}=\bJ+b_0\bB+{\vec b}\times\bE,\\
\label{csm3}
&&{\vec\nabla}\cdot\bB=0,\\
\label{csm4}
&&{\vec\nabla}\times\bE+\frac{\pt \bB}{\pt t}=0,
\end{eqnarray}
where $J^0$ and $\vec J$ are source charge and source current (we assume the ordinary dielectric constant and magnetic permeability to be unit), respectively, and we have introduced the abbreviation $b^\m=(b_0,\vec b)$ with $b_0=e^2\D\m/(4\p^2)$ and $\vec b=e^2\D\vec p/(4\p^2)$. This set of equations constitutes the basis of the Maxwell-Chern-Simons (MCS) electrodynamics. Note that if we identify $b^\m$ with the gradient of a background pseudoscalar field $\h(x)$ (which is called the axion field), $b_0=\pt_t\h$ and $\vec b=\vec\nabla\h$, the above set of equations also describes the axion electrodynamics.

The purpose of the present article is to study the electrodynamics of chiral matter described by the above MCS equations. We will assume the four vector $b^\m$ to be constant. In this situation, the presence of $b^\m$ picks up a direction in spacetime and thus violates the Lorentz invariance; in fact, similar set of equations also appeared in the study of possible Lorentz violation in fundamental interactions~\cite{Carroll:1989vb}. Furthermore, parity symmetry and time-reversal symmetry are also violated by $b_0$ and $\vec b$, respectively. The metric convention is $g_{\m\n}=g^{\m\n}=\diag(1,-1,-1,-1)$.

\section{Electromagnetic wave in chiral matter}\label{sec:wave}
Let us first consider the propagation of electromagnetic (EM) wave in the chiral matter, i.e., let us seek for wave solutions to the sourceless MCS equations ($J^0=0$ and $\vec J=\vec0$). As usual, we substitute the plane-wave ansatz,
\begin{eqnarray}
\vec E=\vec E_0(K) e^{-i\o t+i\vec k\cdot\vec x},
\vec B=\vec B_0(K) e^{-i\o t+i\vec k\cdot\vec x},
\end{eqnarray}
where $K^\m=(\o, \vec k)$ with $\o$ the frequency and $\vec k$ the wave vector.

\subsection{Case 1: $b_0\neq0$ and $\vec b=\vec 0$}\label{sec:case1}
This is the case when only the CME is present. Without loss of generality, we assume $b_0>0$. The dispersion relation of the EM wave is easy to obtain,
\begin{eqnarray}
\label{disp1}
\o^2=k^2\pm b_0 k,
\end{eqnarray}
where $k=|\bk|$. The amplitudes $\vec E_0$ and $\vec B_0$ satisfy
\begin{eqnarray}
\hat{\vec k}\times\bE_0&=&\pm i\bE_0 ,\\
\bB_0&=&\pm i\frac{k}{\o}\bE_0,
\end{eqnarray}
where $\hat{\vec k}=\vec k/k$. Supposing $\hat{\vec k}=\hat{\vec z}$, we have $\bE_0\propto\hat{\vec x}\mp i\hat{\vec y}$; thus the EM wave is circularly polarized with ``$-$" sign and ``$+$" sign corresponding to right-handed and left-handed modes, respectively. Obviously, the two polarized modes have different phase velocities and thus the matter is birefringent and the novel optical phenomena like Pockels effect (which is usually driven by an applied electric field for media with broken parity symmetry; in our case the presence of $b_0$ violates parity symmetry) can occur~\cite{Hecht:2002}.

The dispersion relation (\ref{disp1}) exhibits a striking feature, i.e., when $k<b_0$, the left-handed photon~\footnote{Throughout this paper, we do not distinguish the terms ``EM wave" and ``photon".} is not a propagating mode because its frequency becomes imaginary, $\o=\pm i\sqrt{kb_0-k^2}$. Furthermore, the branch with $\o=i\sqrt{kb_0-k^2}$ ($k<b_0$) is actually unstable, i.e., its wave amplitude grows exponentially with time. This instability (and its variants) has a number of significant consequences in various physical contexts~\cite{Rubakov:1985nk,Joyce:1997uy,Giovannini:1997eg,Field:1998hi,Boyarsky:2011uy,Boyarsky:2012ex,Akamatsu:2013pjd,Manuel:2015zpa,Hirono:2015rla,Tuchin:2016tks,Xia:2016any,Gorbar:2016klv,Pavlovic:2016mxq,Pavlovic:2016gac,Volovik:2016rgb}. For example, it provides a mechanism for the generation of strong long-wavelength magnetic fields (in, e.g., early universe or neutron stars) by converting the axial charge of fermions (characterised by $b_0$ which is proportional to the chiral chemical potential) to the helicity of the magnetic field, ${\cal H}_B=\int d^3\bx \vec A\cdot\vec B$ with $\vec A$ the vector potential for $\vec B$, via the chiral anomaly relation,
\begin{eqnarray}
\frac{d}{dt}(N_R-N_L)=-\frac{e^2}{4\p^2}\frac{d}{dt}{\cal H}_B,
\end{eqnarray}
where $N_{R,L}=V n_{R,L}$ with $V$ the volume and $n_{R,L}$ the number density of right-handed and left-handed fermions.

\subsection{Case 2: $b_0=0$ and $\vec b\neq\vec 0$ }\label{sec:case2}
In this case, the dispersion relation of the EM wave is given by
\begin{eqnarray}
\label{disp2}
\o^2=k^2+\frac{1}{2}b^2\pm\sqrt{(\bk\cdot{\vec b})^2+\frac{1}{4}b^4},
\end{eqnarray}
where $b=|\vec b|$ and ``$+$" and ``$-$" signs generally correspond to right-handed and left-handed elliptically polarized photons (supposing $\vec b\cdot\vec k\leq0$; otherwise, the ``$+$" (``$-$") sign will correspond to left- (right-)handed photon), respectively. Without loss of generality, we can set $\vec b$ to be along $\hat{\vec z}$ direction and then we have
\begin{eqnarray}
\label{disp3}
\o^2=k_x^2+k_y^2+\lb\sqrt{k_z^2+\frac{1}{4}b^2}\pm\frac{b}{2}\rb^2.
\end{eqnarray}
Obviously, $\o^2$ is always larger than or equal to zero so that both right-handed and left-handed photons are propagating modes and no instability is developed.

We list some interesting consequences of the dispersion relation (\ref{disp3}).\\

(1) The dispersion relation (\ref{disp3}) effectively describes one massive degree of freedom with mass $b$, corresponding to the right-handed wave, and one massless degree of freedom, corresponding to the left-handed wave. This is easily seen if we expand \eq{disp3} at small momentum, $k\ll b$,
\begin{eqnarray}
\label{disp22}
\o^2_+&=&k_x^2+k_y^2+2k_z^2+b^2,\\
\label{disp221}
\o^2_-&=&k_x^2+k_y^2+\frac{k_z^4}{b^2}.
\end{eqnarray}
Further more, the left-handed photon behaves like a nonrelativistic particle when propagating along the direction of $\vec b$; similar dispersion relation of photon also occurs at high density nuclear and quark matter \cite{Yamamoto:2015maz}.The very different dispersion relations of right-handed and left-handed photons may be useful in detecting typical properties of Weyl semimetal by using polarized laser. For example, as the right-handed mode is massive, when it travels with frequency smaller than its mass it will be attenuated over a length of order $1/\sqrt{b^2-\o^2}$; this may be used to measure the parameter $\vec b$. In the following discussions we will mainly consider the case with $\o > b$ so that both modes are propagating.\\

(2) The chiral matter with nonzero $\vec b$ has unusual optical properties. Noticeably, it is intrinsically birefringent, i.e., the refractive indices or the phase velocities of the right-handed and left-handed photons are different. The birefringence can bring novel optical phenomena, like the Faraday effect which is the optical gyrotropy in media with broken time-reversal symmetry (in our case, the $\vec b$ breaks the time reversal symmetry). The Faraday effect in general represents the phase difference, $\D\f(L)=\f_R-\f_L$, between right-handed and left-handed photons as they travel a distance $L$ in the matter. Suppose the incident photon is of frequency $\o$ in direction $\hat{\vec n}$. Then after travelling a distance $L$ along $\hat{\vec n}$, the phase difference would be $\D\f(L)=L[\hat{\vec n}\cdot\vec{k}_L(\o)-\hat{\vec n}\cdot\vec{k}_R(\o)]$. For example, for forward propagating photons (along the incident direction), from \eq{disp3}, we find that if $\hat{\vec n}$ is along $x$ direction (or equivalently, $y$ direction), $\D\f(L)=\o L(1-\sqrt{1-b^2/\o^2})$ ($\o>b$); while if $\hat{\vec n}$ is along $z$ direction, $\D\f(L)=\o L(\sqrt{1+b/\o}-\sqrt{1-b/\o})$ ($\o>b$). Thus the Faraday effect depends on both the magnitude and direction of $\vec b$; it may provide a very useful way to determine $\vec b$ parameter in, e.g., Weyl semimetals. Relevant discussions are also given in Refs. ~\cite{Huerta:2014ula,Zhang:2015,Kargarian:2015,Zhong:2015,Zyuzin:2014,Gorbar:2016ygi,Ozaki:2016vwu}.\\

(3) The presence of $\vec b$ breaks the rotational symmetry and makes the chiral matter optically anisotropic. The optics of anisotropic media can be well described by three principal indices of refraction which satisfy the so-called Fresnel's equation. To find the principal indices of refraction, we notice that we can rewrite the MCS equations (\ref{csm1})-(\ref{csm4}) with $b_0=0$ and without source terms in the following form,
\begin{eqnarray}
\label{csm11}
&&{\vec\nabla}\cdot\vec D=0,\\
\label{csm21}
&&{\vec\nabla}\times\bB+i\o\vec D=0,
\end{eqnarray}
with other two equations unchanged, where we have replaced $\pt/\pt t$ by its Fourier variable $-i\o$. In these equations $\vec D$ is the displacement field which is expressed by
\begin{eqnarray}
\vec D=\bE+\frac{i}{\o}\vec b\times\bE\equiv\vec\e(\o)\vec E,
\end{eqnarray}
where $\vec\e(\o)$ is the frequency-dependent dielectric tensor,
\begin{eqnarray}
\vec\e(\o)=\begin{pmatrix}
1 & -ib/\o &0 \\
ib/\o & 1 &0 \\
0 & 0 &1
\end{pmatrix}.
\end{eqnarray}
Thus the chiral matter of this type is gyrotropic even without applying an external magnetic field and the gyration vector is given by $\vec b/\o$~\cite{Hecht:2002}.
By diagonalizing $\vec\e$, we obtain the three principal indices of refraction (for $\o>b$),
\begin{eqnarray}
n_1&=&\sqrt{1+b/\o},\non
n_2&=&\sqrt{1-b/\o},\non
n_3&=&1,
\end{eqnarray}
which are the three eigenvalues of $\sqrt{\vec\e}$, and the corresponding principal dielectric directions,
\begin{eqnarray}
\vec e_1&=&\frac{1}{\sqrt{2}}(\hat{\vec x}+i\hat{\vec y}),\non
\vec e_2&=&\frac{1}{\sqrt{2}}(\hat{\vec y}+i\hat{\vec x}),\non
\vec e_3&=&\hat{\vec z}.
\end{eqnarray}
Since the three indices of refraction are unequal, the chiral matter with a nonzero $\vec b$ is an optical ``biaxial crystal"; and since the principal dielectric directions are complex, the photon is in general expected to be elliptically polarized.
By eliminating $\vec B$ from \eq{csm4} and \eq{csm21}, and writing in the principal coordinates, we obtain
\begin{eqnarray}
\label{determ}
(k_ik^*_j-k^2\d_{ij}+\o^2n^2_{i}\d_{ij})E_j=0,
\end{eqnarray}
with
\begin{eqnarray}
k_1&=&\frac{1}{\sqrt{2}}\lb k_x-ik_y\rb,\non
k_2&=&\frac{1}{\sqrt{2}}\lb k_y-ik_x\rb,\non
k_3&=&k_z.
\end{eqnarray}
The existence of nontrivial solution of \eq{determ} requires the determinant of the matrix in front of $\vec E$ to vanish, which after some algebra gives
\begin{eqnarray}
\label{eqfresnel}
\sum_{i=1}^3\frac{s_i^* s_i}{n^2-n_i^2}=\frac{1}{n^2},
\end{eqnarray}
where we have defined $\vec s=\vec k/k$ being the propagating direction of the photon, and $n=k/\o$. Equation (\ref{eqfresnel}) is the Fresnel's equation of wave normals which defines a surface in the momentum space known as the normal surface~\cite{Born}. For a given propagating direction specified by $\vec s$, there are in general two solutions for $n^2$ which specify two phase velocities $\o/k=1/n$ of the photon propagating in that direction, corresponding to two independent polarizations of the photon~\footnote{However, in the case that $\vec s$ is along one principal axis, the Fresnel's equation contains singularities and one needs to solve \eq{determ} directly}. The directions of the electric field corresponding to the two polarizations are given by
\begin{eqnarray}
\vec E\propto\begin{pmatrix}
\displaystyle\frac{s_1}{n^2-n_1^2}\\
\displaystyle\frac{s_2}{n^2-n_2^2} \\
\displaystyle\frac{s_3}{n^2-n_3^2}
\end{pmatrix}
\end{eqnarray}
in the principal coordinate. Once we know the Fresnel's equation, a number of optical properties of the matter can be derived, see, e.g., Ref.~\cite{Born}.

\subsection{Case 3: $b_0\neq 0$ and $\vec b\neq\vec 0$ }\label{sec:case3}
This is the most general case when the CME, the anomalous Hall effect, and anomalous charge density are all present. In this case, the dispersion relation of the EM wave is given by the following equation
\begin{eqnarray}
\label{disp31}
(K^2)^2+K^2\tilde{b}^2-(\tilde{b}\cdot K)^2=0,
\end{eqnarray}
where $K^\m=(\o,\vec k)$ and $\tilde{b}^\m=(b_0,-\vec b)$ is the parity-flipped $b^\m$. Although \eq{disp31} can be solved analytically, we will not present the solution $\o=\o(\bk)$ as the expression is long and less transparent. An easier way to extract the physics in this case is to notice that the MCS equations are actually Lorentz covariant if we promote $b^\m$ to be a Lorentz four vector; thus two different $b^\m$'s connected by a Lorentz transformation can be considered as equivalent. Hence, a timelike $b^\m$, i.e., $b^2>0$, is equivalent to the case 1 while a spacelike $b^\m$ (i.e., $b^2<0$) is equivalent to the case 2.
The only case that we have not discussed is the lightlike $b^\m$, with which \eq{disp31} becomes
\begin{eqnarray}
\label{disp32}
(K\pm \tilde{b}/2)^2=0.
\end{eqnarray}
Choosing a representative $b^\m=(b,0,0,-b)$, the dispersion relation of photon is given by
\begin{eqnarray}
\label{disp4}
\o=s\big|\vec k+a\frac{\vec b}{2}\big|-a\frac{b}{2},
\end{eqnarray}
where $a=\pm$ represents the two polarizations and $s=\pm$ represents forward and backward moving modes. The solution (\ref{disp4}) displays no instability.

\section{Electrostatics and magnetostatics in chiral matter}\label{sec:statics}
\subsection{Sourceless magnetic and electric fields}\label{sec:statics:1}
We now turn to discuss the time independent features of the MCS electrodynamics, i.e, the electrostatics and magnetostatics of chiral matter. We begin with the sourceless case, i.e., $J_0=0$ and $\vec J=\vec 0$. Very unlike the normal Maxwell electrodynamics which permit only trivial static $\vec E$ and $\vec B$ fields if no charge or current is present, we will see that the static MCS equations permits nontrivial solutions even in the sourceless case. The MCS equations now read
\begin{eqnarray}
\label{csm1s}
&&{\vec\nabla}\cdot\bE=-{\vec b}\cdot\bB,\\
\label{csm2s}
&&{\vec\nabla}\times\bB=b_0\bB+{\vec b}\times\bE,\\
\label{csm3s}
&&{\vec\nabla}\cdot\bB=0,\\
\label{csm4s}
&&{\vec\nabla}\times\bE=0.
\end{eqnarray}

We focus on two special cases.

{\bf Case 1: $b_0\neq 0$ and $\vec b=\vec 0$.} This is the case with only CME presented. In this case, the solution of electric field from \eqs{csm1s}{csm4s} is trivial, so we set $\vec E=\vec 0$; while the $\vec B$ field satisfies
\begin{eqnarray}
\label{beltrami}
&&{\vec\nabla}\times\bB=b_0\bB,
\end{eqnarray}
which implies that the magnetic field must be a Beltrami field\footnote{In 3 spatial dimensions, a vector field that is parallel to its own curl is called a Beltrami field}. In magnetohydrodynamics, a magnetic field that satisfies \eq{beltrami} is also called a Chandrasekhar-Kendall field or a ``force-free" field as in that case the Lorentz force vanishes and thus the magnetic field does not change the material motion~\cite{Chandrasekhar:1957}. The properties of the solution to \eq{beltrami} in the context of usual magnetohydrodynamics were thoroughly studied, for example, it was shown that for fixed magnetic helicity ${\cal H}_B$, the magnetic field configuration satisfying \eq{beltrami} minimizes the magnetic energy ${\cal E}_B=(1/2)\int d^3\bx \vec B^2$~\cite{Woltjer:1958}.

\begin{figure}[!t]
\begin{center}
\includegraphics[width=7.0cm]{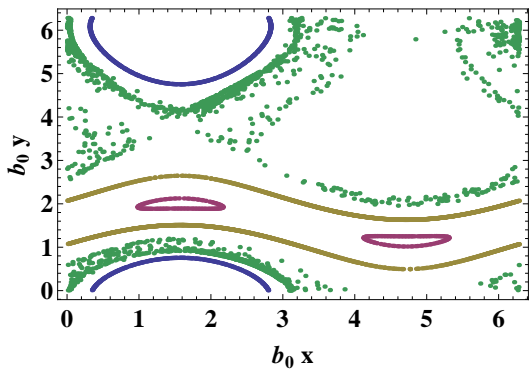}
\caption{A typical Poincare section at $z=0$ for ABC field  with $A^2=1/3, B^2=1, C^2=2/3$. Shown are three ordered (colored by blue, yellow, and purple) and one chaotic (colored by green) magnetic lines.}
\label{poincare}
\end{center}
\end{figure}

Although the static problem (\ref{beltrami}) looks simple, it can exhibit very complex magnetic field configurations according to different boundary conditions. One example with periodic boundary condition is given by
\begin{eqnarray}
\label{abcfield}
B_x&=&B_0\ls A \sin(b_0 z) + C \cos(b_0 y)\rs,\\
B_y&=&B_0\ls B \sin(b_0 x) + A \cos(b_0 z)\rs,\\
B_z&=&B_0\ls C \sin(b_0 y) + B \cos(b_0 x)\rs,
\end{eqnarray}
where $A, B, C$ are three arbitrary constants, and the boundary condition is periodic in $x$, $y$, and $z$ direction so that $\vec B(\vec x+2\p (n_x/b_0)\hat{\vec x}+2\p (n_y/b_0)\hat{\vec y}+2\p (n_z/b_0)\hat{\vec z})=\vec B(\vec x)$ with $n_x, n_y, n_z$ integers. We will call such a magnetic field the Arnold-Beltrami-Childress (ABC) field as it is the magnetic counterpart of the ABC flow in hydrodynamics which is the steady solution of the incompressible Euler equation and satisfies $\bv\propto \vec\o=\vec\nabla\times\bv$~\cite{Arnold:1965,Childress:1970}. It is easy to check that the ABC field saturates the magnetic energy so that ${\cal E}_B=b_0{\cal H}_B/2$. In hydrodynamics, the ABC flow was well studied and was found to be integrable if one of the three parameters $A, B, C$ vanishes; while when $ABC\neq 0$ the flow shows chaotic trajectories~\cite{Dombre:1986}. Similar chaotic structure can also happen for our magnetic lines. In \fig{poincare}, we plot the intersections of four magnetic lines~\footnote{The magnetic line is defined by the equation $d\vec x(s)/d s=\vec B(\vec x)/B_0$ with $s$ parameterizing the line length.} with the planes at $b_0z=0$ mod $2\p$, a plot called Poincare map or Poincare section, where it clearly shows three ordered (marked by blue, purple, and yellow points) and one chaotic (marked by green points) magnetic lines. The ordered magnetic lines form magnetic surfaces while the chaotic magnetic line is dense in the space outside the ordered region. The ordered and chaotic regions are separated by the largest magnetic surfaces which form the boundaries of the chaotic region.

Although we consider only the static configuration in this section, it is worth mentioning that Beltrami configuration is the equilibrium configuration toward which the chiral system evolves through the chiral-anomaly induced inverse cascade of magnetic helicity, according to recent studies in Refs.~\cite{Hirono:2015rla,Yamamoto:2016xtu,Gorbar:2016klv,Xia:2016any,Hirono:2016jps}.

{\bf Case 2: $b_0= 0$ and $\vec b\neq\vec 0$.} In this case, the magnetic and electric fields are coupled to each other, and the presence of one field commonly implies the presence of another one. For example, if a constant magnetic field along $\vec b$ (which is assumed to be along $\hat{\vec z}$ direction) is presented, $\bB=B_0\hat{\vec z}$, then there must be an accompanying electric field distributed in the chiral matter as $\vec E=-B_0z\vec b$ up to a constant. But the opposite is not true, that is, the presence of a constant electric field along $\vec b$ direction induces no magnetic field. Another interesting solution is given by
\begin{eqnarray}
\bB&=&\l\vec b,\\
\bE&=&-\vec\nabla\l,
\end{eqnarray}
where $\l=\l(x,y)$ depends only on $x$ and $y$ and satisfies a Laplace eighen equation with the ``wrong sign",
\begin{eqnarray}
\nabla^2\l=b^2\l,
\end{eqnarray}
which corresponds to the dispersion relation (\ref{disp3}) for right-handed mode at static limit $\o^2\ra 0$. This solution permits solitonic structure around some defects in the chiral matter. For example, putting a charged plane with surface charge density $\s$ at $x=0$, we have $\l(x)=(\s/2b)e^{-b|x|}$ for $b\neq0$ and $x\neq0$~\footnote{When $b=0$, the solution is $\l(x)=-\s|x|/2$, and thus $B_z=0$ and $E_x=(\s/2){\rm sgn}(x)$.}; thus the EM fields read $B_z=(\s/2)e^{-b|x|}$ and $E_x=(\s/2){\rm sgn}(x)e^{-b|x|}$ for $x\neq0$ while other components are zero. This solution means that a charged plane in the chiral matter induces both electric and magnetic fields in such a way that they are orthogonal and both decay exponentially away from the plane. The exponential decay of the fields away from the charge reflects the fact that the photon with dispersion relation (\ref{disp22}) is massive with mass $b$.

\subsection{Point charge and static current}\label{sec:statics:2}
We now turn to discuss the consequences of the static MCS equations with source terms, that is, we want to study the static EM fields induced by charge currents $J^\m$ in chiral matter. For this purpose, we rewrite the MCS equations in the formally covariant form
\begin{eqnarray}
\label{mcscova}
\ls g^{\m\n}\pt^2-\pt^\m\pt^\n-\e^{\m\n\r\s}\tilde{b}_\r\pt_\s\rs A_\n=J^\m,
\end{eqnarray}
where $\tilde{b}^\m=(b_0,-\vec b)$ and the gauge potential $A^\m$ is related to the EM field through the usual definition $F^{\m\n}=\pt^\m A^\n-\pt^\n A^\m$. The solution of this equation can be expressed by using the Green's function,
\begin{eqnarray}
\label{solmcs}
A^\m(K)=G^{\m\n}(K)J_\n(K)+A^\m_0(K),
\end{eqnarray}
where we have expressed the solution in momentum space and $A^\m_0(K)$ is the photon field which satisfies the sourceless MCS equations and has been studied in last subsection. In this subsection, we focus on the photon field generated by $J^\m$, so we will not consider the $A^\m_0$. The Green's function $G_{\m\n}(K)$ is given by
\begin{eqnarray}
G_{\m\n}(K)=-\frac{K^2 g_{\m\n}+i\e_{\m\n\r\s}\tilde{b}^\r K^\s+\tilde{b}_\m \tilde{b}_\n}{(K^2)^2+\tilde{b}^2K^2-(\tilde{b}\cdot K)^2},
\end{eqnarray}
which can be checked by directly substituting \eq{solmcs} in to \eq{mcscova} and neglecting terms due to charge conservation $K_\m J^\m=0$~\cite{Lehnert:2004hq}.

In the static limit, $\o\ra0$, the Green's function is
\begin{eqnarray}
G_{\m\n}(\o=0,\bk)=\frac{\bk^2 g_{\m\n}-i\e_{\m\n\r i}\tilde{b}^\r k^i-\tilde{b}_\m \tilde{b}_\n}{(\bk^2)^2-\tilde{b}^2\bk^2-(\vec b\cdot \bk)^2}.
\end{eqnarray}
We notice that at large $\bk$, $k\gg b_0, b$, the above Green's function becomes the usual one without any anomalous effects which means the short-distance static EM field induced by a charge current $J^\m(\vec x)$ is the usual Maxwell fields. So in the following we focus on large-distance fields.

{\bf Case 1: $b_0\neq 0$ and $\vec b=\vec 0$.} First, let us consider the case with only CME present. In this case, the components of the Green's function read
\begin{eqnarray}
G_{00}(\bk)&=&\frac{1}{\bk^2},\\
G_{0i}(\bk)&=&G_{i0}(\bk)=0,\\
G_{ij}(\bk)&=&-\frac{\d_{ij}}{\bk^2-b_0^2}+\frac{i\e_{ijl}k^l}{b_0(\bk^2-b_0^2)}-\frac{i\e_{ijl}k^l}{b_0\bk^2}.
\end{eqnarray}
Thus, a point charge will still generate the usual Coulomb electric field, but the electric current can generate very different magnetic field. This is most easily seen in coordinate space in which the Green's function $G_{ij}(\bx)$ reads
\begin{eqnarray}
\label{green}
G_{ij}(\bx)&=&-\frac{\d_{ij}\cos(b_0|\bx|)}{4\p|\bx|}\non&&
-\frac{\e_{ijl}x^l}{4\p b_0|\bx|^3}\ls b_0 |\bx|\sin(b_0|\bx|)+\cos(b_0|\bx|)\rs\non
&&+\frac{\e_{ijl}x^l}{4\p b_0 |\bx|^3},
\end{eqnarray}
which is very different from the familiar $\d_{ij}/|\bx|$ form for ordinary electromagnetism. The first two terms are oscillatory which is due to the real poles $\pm b_0$ in $G_{ij}(\bk)$.

\begin{figure}[!t]
\begin{center}
\includegraphics[width=5.0cm]{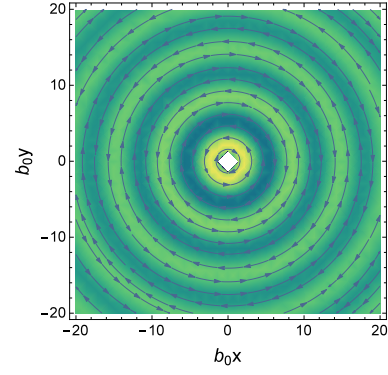}
\caption{Illustration of the magnetic field generated by an infinitely long straight wire along $z$-axis. The arrows represent the transverse magnetic lines and the density plot represents the magnitude of the $z$-component of the magnetic field.}
\label{mag_b0}
\end{center}
\end{figure}
\begin{figure}[!t]
\begin{center}
\includegraphics[width=4.0cm]{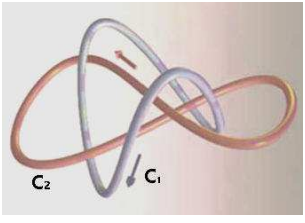}
\caption{Illustration of two linked loops $C_1$ and $C_2$ carrying constant electric currents $I_1$ and $I_2$, respectively.}
\label{twoloops}
\end{center}
\end{figure}

The vector potential $\vec A(\bx)$ generated by a current $\vec J(\bx)$ is then given by
\begin{eqnarray}
\label{vectorpotential}
A^i(\bx)=\int d^3\by G^{ij}(\bx-\by)J_j(\by).
\end{eqnarray}
One interesting feature of this vector potential is that the magnetic field generated by an electric current can have component parallel to the current, see illustration in \fig{mag_b0} in which we present the spacial distribution of the magnetic field generated by an infinitely long straight wire. An intriguing consequence of \eq{green} is that the interaction between two charge-current loops contains a topological component due to the third term in \eq{green} which is independent of the shapes and relative distance between the two loops and this topological component becomes important when condition $L\gg 1/b_0$ ($L$ is the shortest distance between the two loops) is satisfied, as first noticed in Ref.~\cite{Khaidukov:2013sja}. To see this, we write down the interaction energy between two static classical currents $J^i_1(\bx)=I_1\oint_{C_1}\d^{(3)}(\bx-\bx_1) dx_1^i$ and $J^i_2(\bx)=I_2\oint_{C_2}\d^{(3)}(\bx-\bx_2) dx_2^i$, where $I_{1,2}$ are the strengths of the currents and $C_{1,2}$ denote the two loops, see \fig{twoloops},
\begin{eqnarray}
\d E&=&\int d^3\bx d^3\by J^i_{1}(\bx)G_{ij}(\bx-\by)J^j_{2}(\by)\non
&=&\frac{I_1I_2}{b_0}\oint_{C_1}\oint_{C_2} dx_1^i dx_2^j\frac{\e_{ijl}(x_1-x_2)^l}{4\p |\bx_1-\bx_2|^3}\non
&=&\frac{I_1I_2}{b_0}{\cal L}(C_1, C_2),
\end{eqnarray}
where ${\cal L}(C_1, C_2)$ is the Gauss linking number of loops $C_1$ and $C_2$ which is topological invariant that measures the degree of linkages between $C_1$ and $C_2$. Similarly, one can write down the interaction energy due to the first two terms in $G_{ij}(\bx)$ which are not topological and are important when the two loops are not separated too far away.

{\bf Case 2: $b_0= 0$ and $\vec b\neq\vec 0$.} In this case, the components of the Green's function read
\begin{eqnarray}
\label{g00}
G_{00}(\bk)&=&\frac{\bk^2}{(\bk^2)^2+b^2\bk^2-(\vec b\cdot\vec k)^2},\\
\label{g0i}
G_{0i}(\bk)&=&-G_{i0}(\bk)=-\frac{i\e_{ijl}b^j k^l}{(\bk^2)^2+b^2\bk^2-(\vec b\cdot\vec k)^2},\\
\label{gij}
G_{ij}(\bk)&=&-\frac{\bk^2\d_{ij}+b_ib_j}{(\bk^2)^2+b^2\bk^2-(\vec b\cdot\vec k)^2}.
\end{eqnarray}
Without loss of generality, we assume $\vec b$ to be along $\hat{\vec z}$ direction. The corresponding Green's functions in coordinate space are
\begin{eqnarray}
\label{g00c}
G_{00}(\bx)&=&-\nabla^2 f(\bx),\\
\label{g0ic}
G_{0i}(\bx)&=&-G_{i0}(\bx)=\e_{ijl}b^j \pt^l f(\bx),\\
\label{gijc}
G_{ij}(\bx)&=&(\d_{ij}\nabla^2-b_ib_j)f(\bx),
\end{eqnarray}
where
\begin{eqnarray}
\label{funcf}
f(\bx)&=&\int\frac{d^3\bk}{(2\p)^3}\frac{e^{i\bk\cdot\bx}}{(\bk^2)^2+b^2\bk^2-(\vec b\cdot\vec k)^2}\non
&=&\frac{1}{8\p b}\int_0^\infty dk_\perp \frac{J_0(k_\perp b|\bx_\perp|)}{\sqrt{k_\perp}}\non&&\times\lb\frac{e^{-b\sqrt{k_\perp^2-ik_\perp}|z|}}{i\sqrt{k_\perp-i}}-\frac{e^{-b\sqrt{k_\perp^2+ik_\perp}|z|}}{i\sqrt{k_\perp+i}}\rb,
\end{eqnarray}
where $J_0(x)$ is the Bessel function, and $\bx_\perp$ and $\vec k_\perp$ stand for the coordinate and momentum transverse to $\vec b$.

\begin{figure}[!t]
\begin{center}
\includegraphics[width=4.0cm]{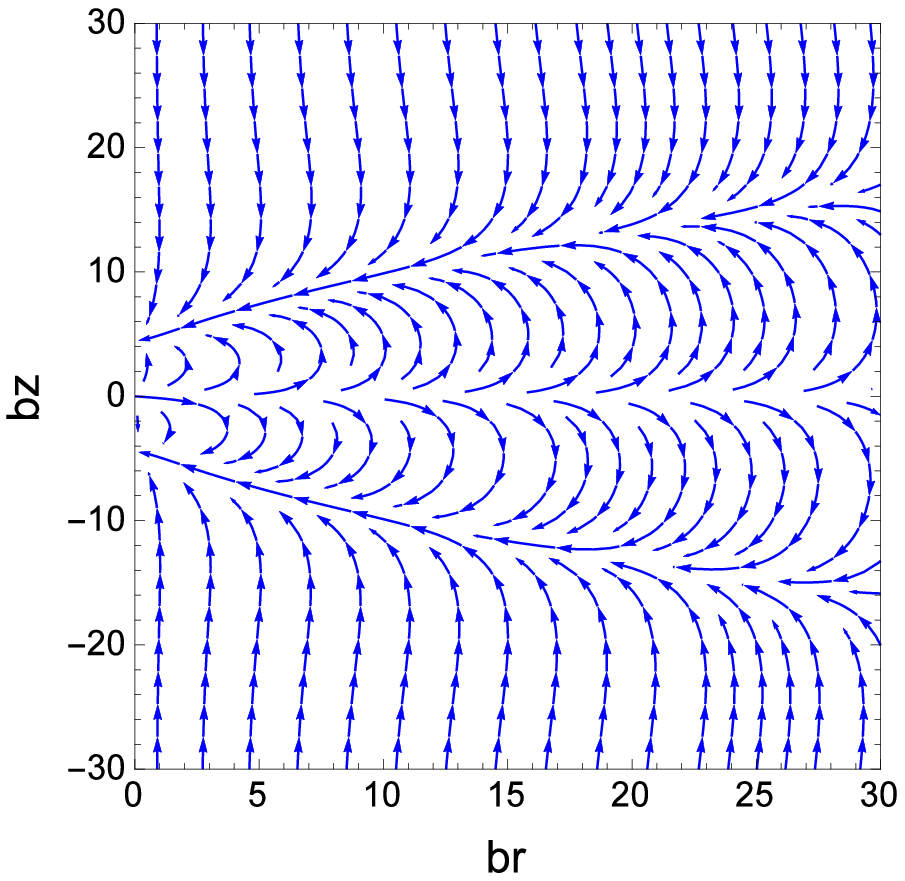}
\includegraphics[width=4.0cm]{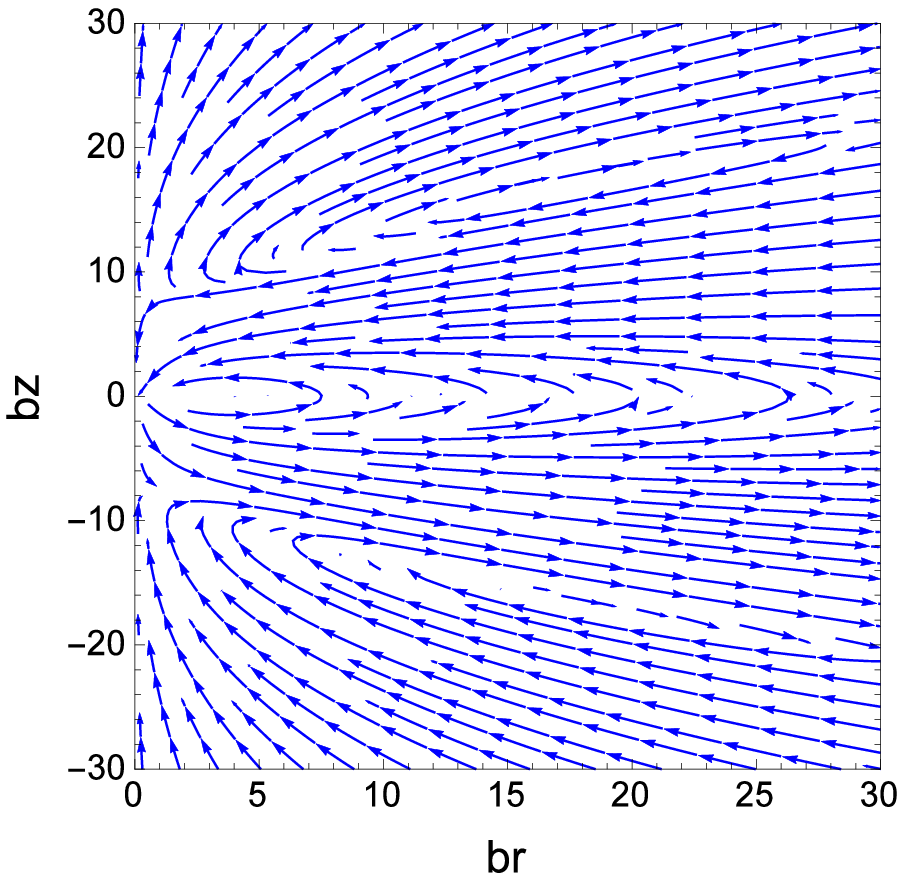}
\caption{Illustration of the electric field (left) and magnetic field (right) generated by a point charge at the origin. The vector $\vec b$ is along $\hat{\vec z}$ direction, $r$ is the perpendicular distance from the $z$-axis.}
\label{ele_b1}
\end{center}
\end{figure}

\begin{figure}[!t]
\begin{center}
\includegraphics[width=4.0cm]{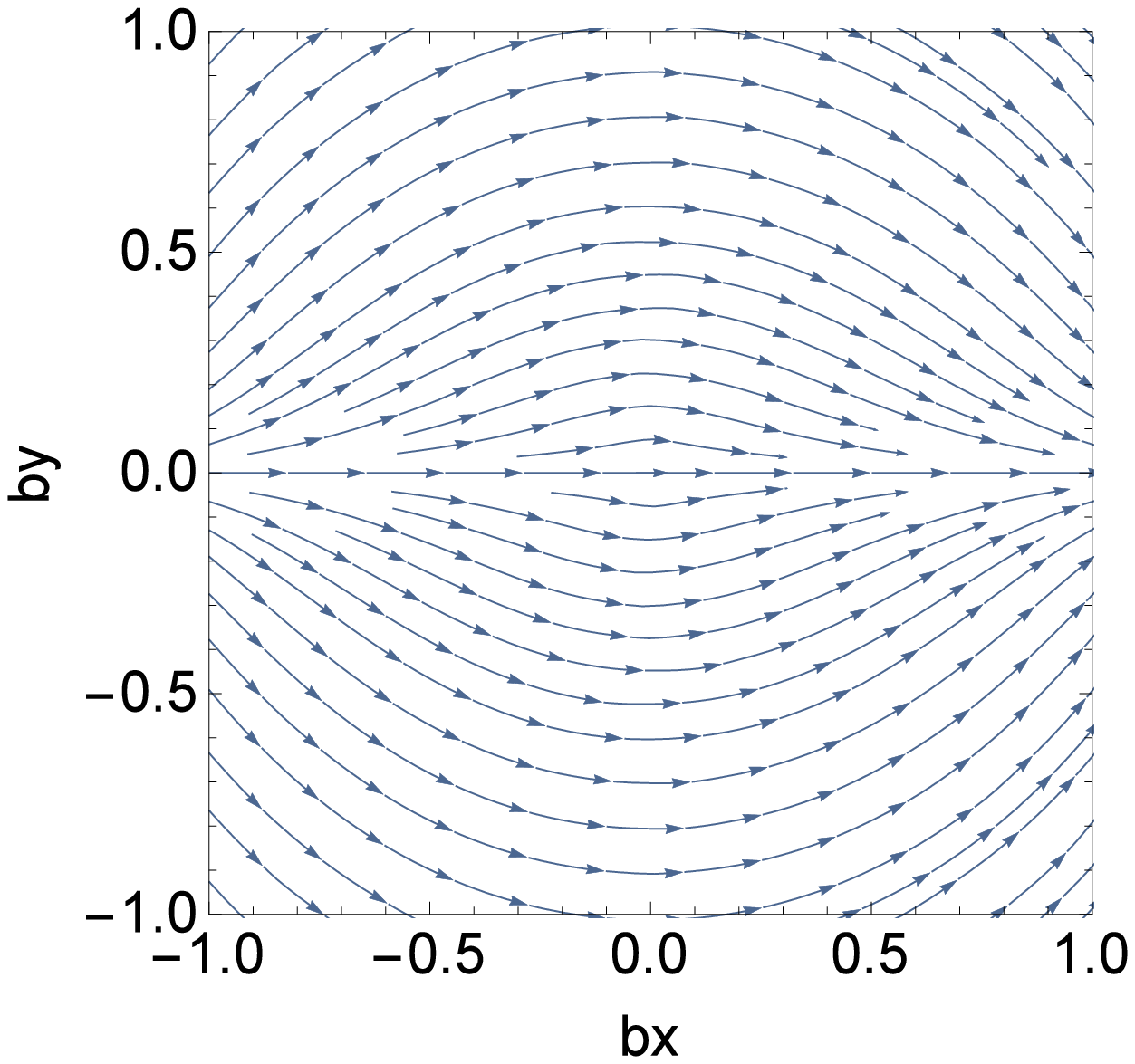}
\includegraphics[width=4.0cm]{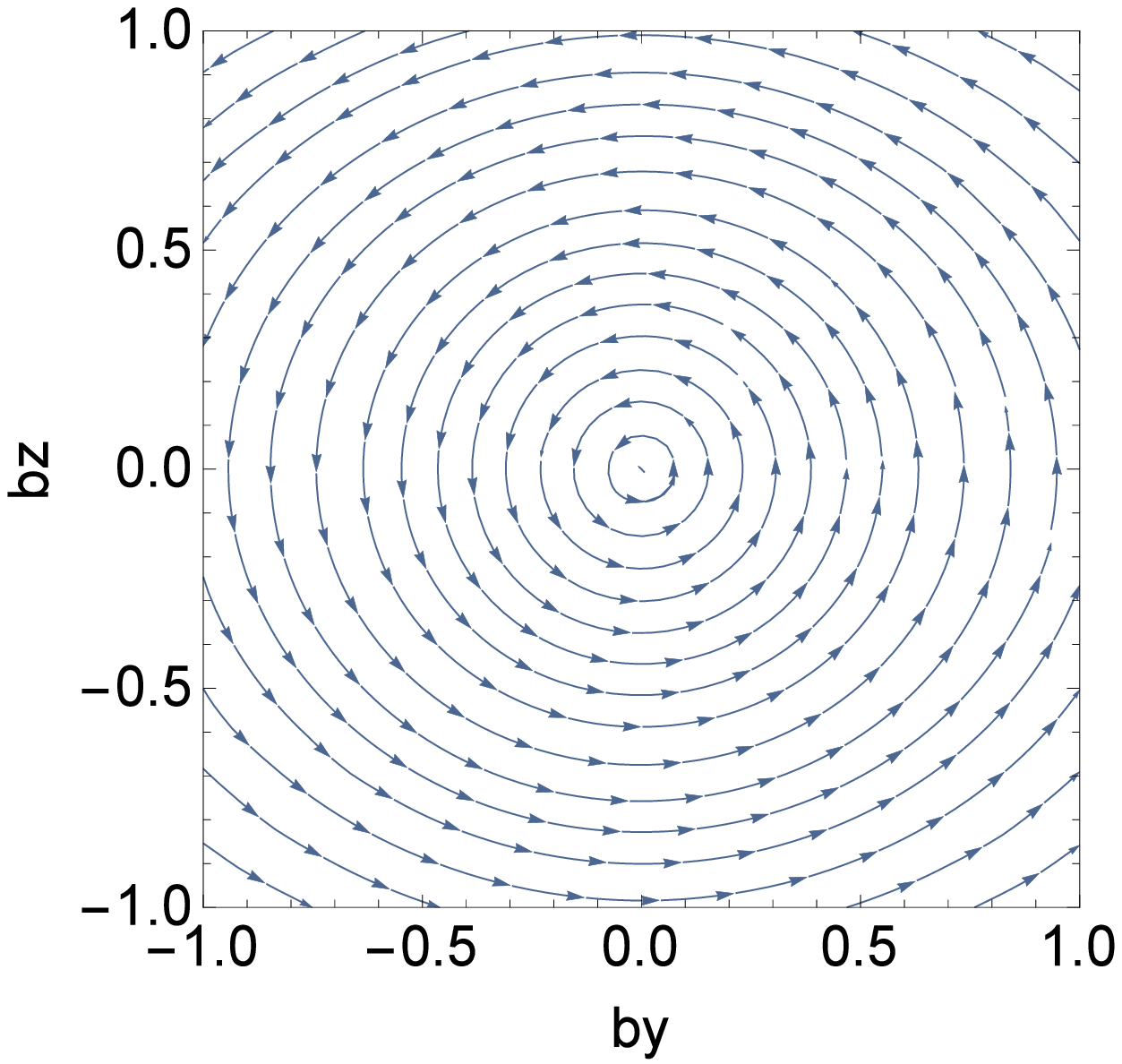}
\caption{Illustration of the electric field in the $x-y$ plane (left) and magnetic field in the $y-z$ plane (right) generated by an infinitely long straight wire along $x$-axis. The vector $\vec b$ is along $\hat{\vec z}$ direction. }
\label{ele_b2}
\end{center}
\end{figure}

The most striking feature here is that $G_{i0}$ and $G_{0i}$ do not vanish which means a point charge can generate a static magnetic field and an electric current can generate a static electric field. In \fig{ele_b1}, we illustrate the electric field and the magnetic field generated by a point charge, and in \fig{ele_b2}, we illustrate the electric field and the magnetic field generated by an infinitely long straight wire along $x$-axis. It is interesting to notice that the appearance of $\vec b$ leads to quite novel electric and magnetic fields distribution. Note that if the infinitely long straight wire is parallel to the vector $\vec b$, it does not generate any electric field and it generates a magnetic field identical to the case with $b=0$.

\section{Summary}\label{sec:summary}
In summary, we have studied the electrodynamics and optics of chiral matter based on the Maxwell-Chern-Simons equations which contain effects that violate parity and time reversal symmetries. A number of novel electromagnetic (EM) and optical properties of chiral matter are explored.

For chiral matter characterized by a nonzero $b_0$ (i.e., when only the chiral magnetic effect is present): (1) The propagation of the left-handed EM wave with wavelength larger than $1/b_0$ is unstable. (2) The sourceless static magnetic field is a Beltrami field and in general there can exist chaotic magnetic lines. (3) The magnetic field of an electric current can have component that is parallel to the current. The interaction between two far-separated static current loops is topological.

For chiral matter characterized by a nonzero vector $\vec b$ (i.e., when the anomalous Hall effect and anomalous charge density are present): (1) The propagation of the EM wave is always stable and causal. (2) The right-handed photon is massive while the left-handed photon has a nonrelativistic dispersion relation when propagating along the direction of $\vec b$. (3) Such chiral matter is generally birefringent and actually an optical ``biaxial crystal" with three different principal refractive indices. We derived the corresponding Fresnel's equation for such chiral matter. (4) There is an intrinsic Faraday effect associated with such chiral matter. (5) In general, a point charge can induce a magnetic field while an electric current can induce an electric field.

Finally, we emphasize that, the present work can be considered to be complementary to the previous studies in, e.g., Refs.~\cite{Kharzeev:2010gd,Huang:2013iia,Jiang:2014ura} where the authors considered the situation that the EM fields are fixed while the matter fields (e.g., chiral density and electric charge density) are fluctuating. In this article, we consider the opposite situation where the matter parameters $b_0$ and $\vec b$ are fixed while the EM fields are varying in the fixed matter background. In a real matter, however, the presence of EM fields would change $b_0$ via the chiral anomaly and also be able to induce other terms in the Maxwell-Chern-Simons equations, e.g., the terms representing the Ohm current and electric charge density. Thus, the fluctuations of the EM fields would be coupled to the fluctuations of the chiral density and electric charge density, and new collective modes would arise. It will be an interesting future task to explore such new phenomena.

\emph{Acknowledgments}---
We thank Y. Hidaka, Y. Hirono, D. Kharzeev, J. Liao, S. Lin for useful discussions. This work is supported by Shanghai Natural Science Foundation with Grant No. 14ZR1403000, 1000 Young Talents Program of China, and  NSFC with Grant No. 11535012 and No. 11675041. G. C. is also supported by China Postdoctoral Science Foundation with Grant No. KLH1512072.

\end{document}